\documentclass[preprint,authoryear]{elsarticle}

\usepackage{amssymb}
\usepackage{graphicx}
\usepackage{subfigure}	        % Sottofigure
\usepackage{aas_macros}
\usepackage{multirow}

\journal{Nuclear Physics A}

\begin{document}

\begin{frontmatter}


\title{Spectral and polarimetric characterization of the Gas Pixel Detector filled with dimethyl
ether}

\author[IASF]{F.~Muleri\corref{cor}}
\cortext[cor]{Corresponding author.}
\ead{fabio.muleri@iasf-roma.inaf.it}
\author[IASF]{P.~Soffitta}
\author[INFN]{L.~Baldini}
\author[INFN]{R.~Bellazzini}
\author[INFN]{A.~Brez}
\author[IASF]{E.~Costa}
\author[IASF,UTOV]{S.~Fabiani}
\author[EPFL]{F.~Krummenacher}
\author[INFN]{L.~Latronico}
\author[IASF]{F.~Lazzarotto}
\author[INFN]{M.~Minuti}
\author[INFN]{M.~Pinchera}
\author[IASF]{A.~Rubini}
\author[INFN]{C.~Sgr{\'o}}
\author[INFN]{G.~Spandre}
\address[IASF]{IASF/INAF, Via del Fosso del Cavaliere 100, I-00133 Roma, Italy}
\address[INFN]{INFN sez. Pisa, Largo B. Pontecorvo 3, I-56127 Pisa, Italy}
\address[UTOV]{Universit\`{a} di Roma ``Tor Vergata'', Dipartimento di Fisica, via della Ricerca
Scientifica 1, I-00133 Roma, Italy}
\address[EPFL]{EPFL, Route Cantonale CH-1015 Lausanne, Switzerland}

\begin{abstract}
The Gas Pixel Detector belongs to the very limited class of gas detectors optimized for the
measurement of X-ray polarization in the emission of astrophysical sources. The choice of the
mixture in which X-ray photons are absorbed and photoelectrons propagate, deeply affects both the
energy range of the instrument and its performance in terms of gain, track dimension and
ultimately, polarimetric sensitivity. Here we present the characterization of the Gas Pixel Detector
with a 1~cm thick cell filled with dimethyl ether (DME) at 0.79~atm, selected among other mixtures
for the very low diffusion coefficient. Almost completely polarized and monochromatic photons were
produced at the calibration facility built at INAF/IASF-Rome exploiting Bragg diffraction at nearly
45~degrees. For the first time ever, we measured the modulation factor and the spectral
capabilities of the instrument at energies as low as 2.0~keV, but also at 2.6~keV, 3.7~keV, 4.0~keV,
5.2~keV and 7.8~keV. These measurements cover almost completely the energy range of the instrument
and allows to compare the sensitivity achieved with that of the standard mixture, composed of helium
and DME.
\end{abstract}


\end{frontmatter}

\section{Introduction}

Detectors able to image charged particle tracks in a gas have been developed over the last few
years for different applications. One of the most promising is the possibility to resolve the path
of photoelectrons emitted in the gas in consequence of a photoelectric absorption. The
reconstruction of the initial direction of photoelectron emission opens the way for measuring the
state of polarization of the absorbed photons because the former is modulated with respect to the
direction of the photon electric field with a $\cos^2$ dependency. This makes the photoelectric
effect a good analyzer of X-ray polarization, and a perfect one for absorption from spherically 
symmetric shells.

Only a few gas detectors can resolve so finely the photoelectron tracks to accurately reconstruct
the initial direction of emission \citep{Bellazzini2006, Black2007}. One of the most sensitive is
the Gas Pixel Detector (GPD hereafter), developed by INFN-Pisa and INAF/IASF-Rome
\citep{Costa2001,Bellazzini2007} and currently inserted in the focal plane of several future
satellite missions \citep{Bellazzini2010b, Costa2010}. The gas cell is 1~cm or 2~cm thick and a
number of mixtures of helium, neon or argon and dimethyl ether (DME hereafter) at 1~atm or 2~atm
have been used, the choice of the gas being of fundamental importance for the polarimetric
performance of the detector. A hard limit to the lower energy threshold of the instrument is about
twice the binding K-shell energy of the absorbing component because above this threshold the
photoelectron track, modulated with polarization, prevails on the isotropic one of the Auger
electron. Photoelectron range is determined by density, while the average atomic number fixes the
mean free path for scatterings with atomic nuclei, which is the length scale on which polarimetric
information is smeared. The diffusion coefficient influences the blurring of the photoelectron track
during drift in the gas cell and eventually the possibility to resolve the initial part of the
photoelectron path and reconstruct correctly the direction of emission. We developed a Monte Carlo
software to easily explore the behavior of the instrument to different mixtures and to subsequently
test a subset of the most interesting ones.

Recently \citet{Muleri2008} measured the modulation factor $\mu$, namely the amplitude of the
response of the instrument for completely polarized photons, for the GPD filled with helium 20\% and
DME 80\% at 2.6~keV, 3.7~keV and 5.2 keV. This data confirmed that measured values are basically
consistent with what is expected on the basis of the Monte Carlo software and proved that X-ray
polarimetry in Astrophysics with the GPD is feasible. In this paper we characterize the behavior of
the GPD with a different gas, i.e. pure DME at 0.79~atm. In particular, we describe the
configuration of the GPD and the calibration sources we used in Section~\ref{sec:Setup}, while
spectral capabilities of the instrument are discussed in Section~\ref{sec:Spectral}. The measurement
of the modulation factor between 2.0~keV and 7.8~keV is reported in Section~\ref{sec:Mod}, together
with the comparison with Monte Carlo results and what was reported previously on the He 20\% and DME
80\% mixture. Note that this is the first time that the modulation factor of a gas polarimeter is
presented at energies as low as 2.0~keV.

\section{Set-up} \label{sec:Setup}

\subsection{Detector configuration}

The Gas Pixel Detector is composed of a sealed 1~cm thick gas cell enclosed by a 50~$\mu$m beryllium
window, a GEM \citep[Gas Electron Multiplier,][]{Sauli1997} which collects and amplifies primary
electrons produced by photoelectrons in the gas cell, and a finely subdivided pixelized detector
\citep{Costa2001,Bellazzini2007}. The last component, based on a VLSI ASIC realized in 0.18~$\mu$m
CMOS technology, is the actual breakthrough of the instrument \citep{Bellazzini2006}, which
otherwise is fundamentally an array of standard yet exceptionally small independent proportional
counters. The top metal layer of the CMOS is fully pixellated to collect the charge produced in the
common gas volume and allows to obtain a true 2D image of the photoelectron track even at low
energy, thanks to the small (50~$\mu$m) pixel size. The acquisition is self-triggered and only a
small window of about a thousand of pixels enclosing the track is actually read-out in place of the
whole matrix. The chip is 15$\times$15~mm$^2$ and comprises 105,600 pixels arranged in a hexagonal
pattern.

The cell is sealed but can be refilled to test different gases and typically mixtures of helium,
neon or argon and DME are used. DME is used to reduce diffusion and also as a quencher, but it
acts as the actual absorber in the case of helium mixtures. The first application of the
instrument in Astrophysics is expected in the 2-10~keV energy range and within this interval the
standard mixture is helium 20\% and DME 80\% \citep{Muleri2008}. This was preferred to mixtures of
neon because of the longer photoelectron path and lower diffusion for equivalent efficiency, which
assure a higher polarimetric sensitivity at low energy where the largest part of photons are
concentrated. In this paper we push the use of low-diffusion mixtures to the extreme, exploring the
use of a pure DME gas at 0.79~atm (0.8~bar). Since helium is basically transparent to X-rays in the
2-10~keV energy range, we expect a sensitivity comparable to the standard mixture, with a possible
enhancement because of the lower diffusion.

An improvement with respect to previous versions of the GPD is the use of a laser-etched GEM
made of liquid crystal polymer which shows a better temporal gain stability \citep{Tamagawa2009}. A
drawback is that the smallest pitch available was only 80~$\mu$m (instead of 50~$\mu$m of previous
detectors), and this has proved to be insufficient to avoid the emergence of systematic effects due
to undersampling of short tracks, discussed and removed, as explained in Section~\ref{sec:Cuts}.
GEMs with smaller pitch are now in production and will be used for the next GPD prototype. Moreover
the thickness is 100~$\mu$m instead of 50~$\mu$m.

The characteristics of the GPD used are summarized in Table~\ref{tab:GPD}.

\begin{table}
\centering
\begin{tabular}{rl}
Area: & 15$\times$15~mm$^2$ \\
Active area fill fraction: & 92\% \\
\hline
Window: & 50~$\mu$m, beryllium \\
Mixture: & DME 100\%, 0.79~atm (0.8~bar) \\
Cell thickness: & 1~cm \\
\hline
GEM material: & copper-coated liquid crystal polymer \\
GEM pitch: & 80 $\mu$m \\
GEM holes diameters: & 48 $\mu$m\\
GEM thickness: & 100~$\mu$m \\
GEM voltages: & V$_{drift}$=3200~V, V$_{top}$=1145~V, V$_{bottom}$=500~V \\
Gain: & 500 \\
\hline
Pixels: & 300$\times$352, hexagonal pattern\\
Pixel noise: & 50 electrons ENC \\
Full-scale linear range: & 30000 electrons \\
\hline
Peaking time: & 3-10 $\mu s$, externally adjustable \\
Trigger mode: & internal, external or self-trigger \\
Self-trigger threshold: & 2000 electrons \\
Pixel trigger mask: & individual \\
\end{tabular}
\caption{Main characteristics of the GPD prototype studied in this paper.}
\label{tab:GPD}
\end{table}

\subsection{Calibration facility}

The GPD was characterized at the X-ray facility built at INAF/IASF-Rome. Although its detailed
description is beyond the scope of this paper, in the following we briefly present what is relevant
to measurements presented below.

Polarized and monochromatic X-rays are produced by Bragg diffraction at 45$^\circ$
\citep{Evans1977}. Incident radiation on a crystal can be decomposed in two components, polarized
parallel ($\pi$-component) and perpendicularly ($\sigma$-component) to the diffraction
plane. The latter is more effectively diffracted because the ratio $k$ between the integrated
reflectivity of the $\pi$ and $\sigma$ components is always smaller than 1. Hence diffracted
radiation is (partially) polarized and the degree of polarization $\mathcal{P}$ is:
\begin{equation}
\mathcal{P} = \frac{1-k}{1+k}. \label{eq:PolarizationDegree}
\end{equation}

\begin{figure}[tbp]
\begin{center}
\subfigure[\label{fig:Bragg_k1}]{\includegraphics[angle=0,totalheight=4.2cm]{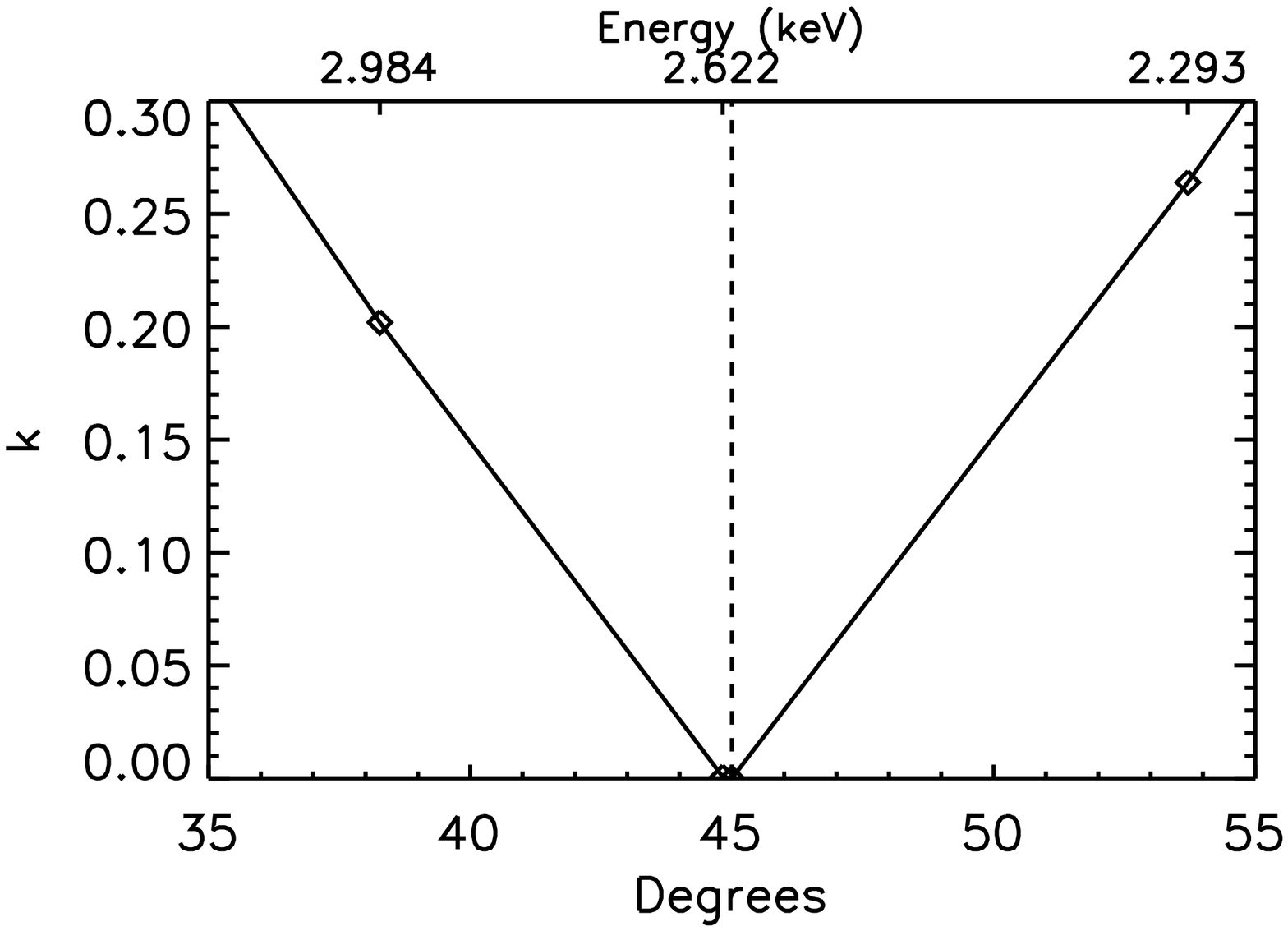}}\hspace{2mm}
\subfigure[\label{fig:Bragg_P1}]{\includegraphics[angle=0,totalheight=4.2cm]{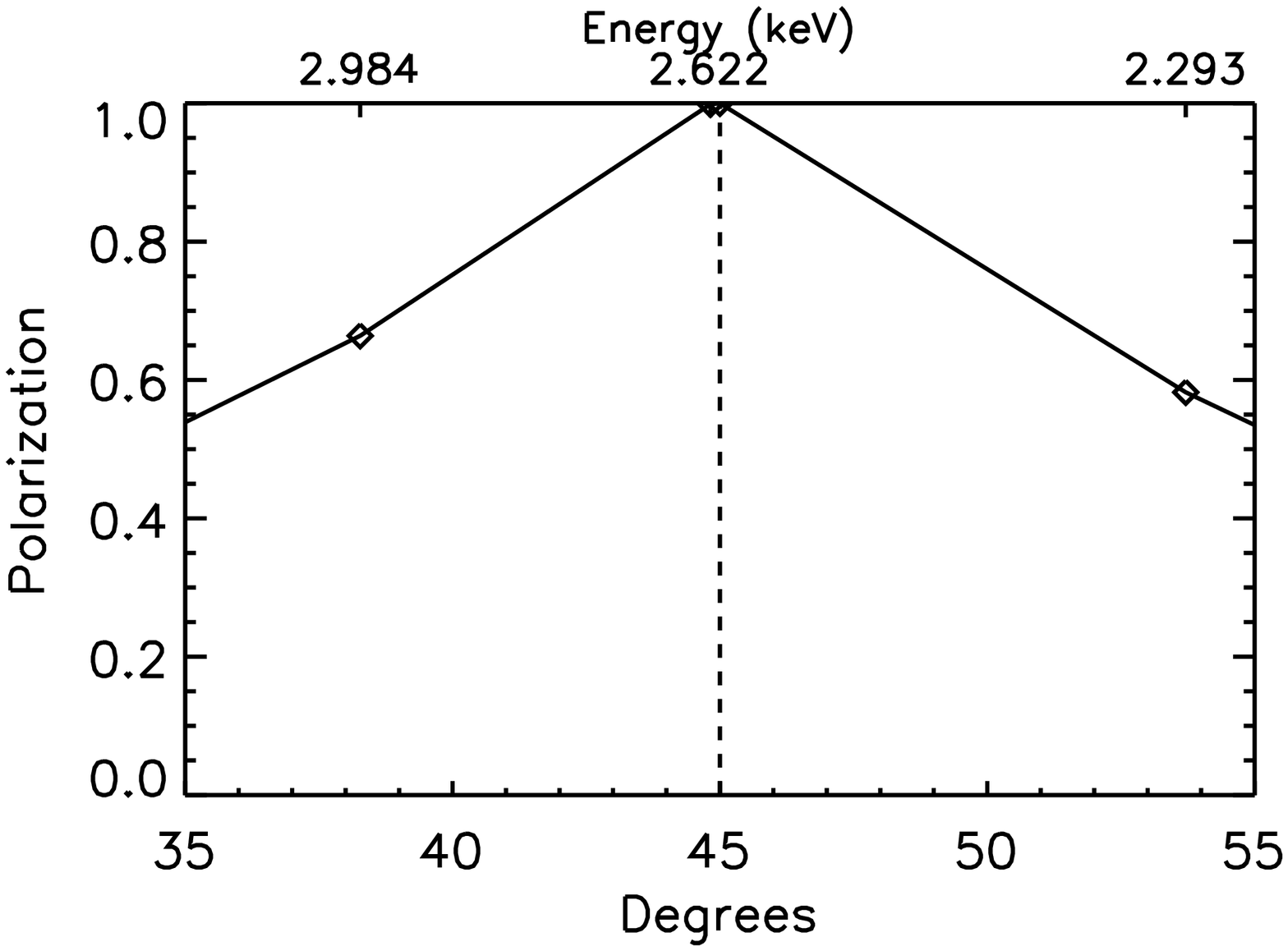}}
\end{center}
\caption{{(\bf a}) Dependence of $k$ with incident angle and energy, related by Bragg's law, for
graphite crystals. The value of $k$ was calculated by \citet{Henke1993} ({\bf b}) Expected degree
of polarization, derived from Equation~(\ref{eq:PolarizationDegree}).}
\end{figure}

If the incident angle $\theta$ is 45$^\circ$, $k=0$ and consequently $\mathcal{P}=1$. For
intermediate values, $k$ can be calculated and the value as a function of $\theta$ is reported in
Figure~\ref{fig:Bragg_k1} for graphite crystals. The large dependence of $k$ on the incident angle
requires the value of $\theta$ to be tightly constrained to prevent the dilution of the average
degree of polarization (cf. Figure~\ref{fig:Bragg_P1}). The angular constraint also selects the
energy of diffracted radiation, related to $\theta$ by Bragg's Law:
\begin{equation}
E(\theta) = \frac{nhc}{2d\sin\theta}, \label{eq:BraggLaw}
\end{equation}
where $h$ and $c$ are respectively Planck's constant and the speed of light, $d$ the crystal
lattice spacing and $n$ an integer which specifies the diffraction order.

We already presented a prototype source based on Bragg diffraction, which exploits lead-glass
capillary plates to constrain to 45$^\circ$ the incident and diffraction angles and small (2~W)
X-ray tubes to produce the radiation to be diffracted \citep{Muleri2007}. This source was used to
generate polarized photons at 2.6~keV, 3.7~keV and 5.2~keV and calibrate at these energies the GPD
filled with a He-DME mixture \citep{Muleri2008}. An aluminum crystal and an X-ray tube with anode
made of calcium were exploited to produce 3.7~keV polarized photons, while 2.6~keV and 5.2~keV were
obtained by first and second order diffraction on graphite of copper X-ray tube radiation. The
former configuration is particularly effective in terms of higher flux and control of the output
state of polarization because K$\alpha$ fluorescence emission of calcium is well in accordance with
Bragg energy at 45$^\circ$ for aluminum. Then (almost) all incident photons have a well-defined
energy, that of K$\alpha$ line of calcium, and are diffracted exactly at the Bragg angle given by
Equation~(\ref{eq:BraggLaw}) and the degree of polarization is precisely calculated with
Equation~(\ref{eq:PolarizationDegree}). A trade-off between flux (low collimation) and high
polarization (high collimation) was instead necessary for diffraction on graphite because X-ray
tubes with anodes in accordance with Bragg energy are not available in this case and continuum
bremsstrahlung emission is to be used.

\begin{figure}
\centering
\includegraphics[width=8cm]{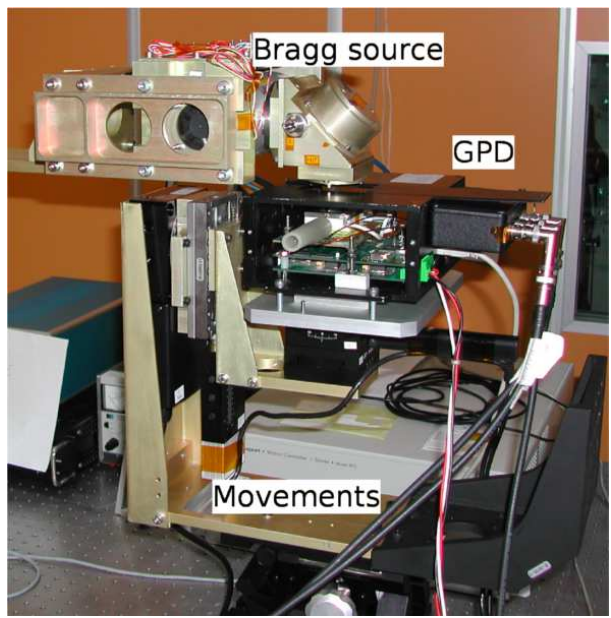}
\caption{Setup of the measurements. X-rays, generated by a 50~W tube manufactured by Oxford
Instruments, are emitted horizontally and diffracted downwards by a crystal oriented at 45$^\circ$.
Two collimators constrain the divergence of the incident and of the diffracted photons and a
diaphragm, placed after the second collimator, limits the size of the beam. The detector is mounted
on a platform which is allowed to rotate, move and incline with respect to the incident beam
with high precision, $<1~\mu$m for movements and $<1$~arcsec for rotation/inclination.}
\label{fig:Setup}
\end{figure}

After the construction of this first prototype, we built a more powerful source based on the same
concept but with some differences. The most important is that the crystal is mounted on a manual
stage which allows two axes tilt regulation in the range $\pm3^\circ$ to achieve the best alignment
to the Bragg condition. More crystals are available to produce radiation at different energies
and, thanks to more powerful X-ray sources (50~W), a tight collimation can be retained even for
diffraction of continuum radiation, which is less effective than the use of line in accordance to
Bragg condition. A mechanical assembly combined with motorized and manual stages, which complete
what we call X-ray facility, allows the detector to be moved, rotated and inclined with respect to
the beam (see Figure~\ref{fig:Setup}).

Within the work presented in this paper, we used three different crystals, PET, graphite (grade D,
1.2$^\circ$ of mosaic spread) and aluminum. In the first two cases, we used the improved version of
the Bragg source and exploited the diffraction of continuum photons produced by an X-ray tube with
anode made of titanium, which has the thinnest window among our medium power tubes. We constrained
both incident and diffraction angles on the crystal with two capillary plates which provide a
collimation $1/40$ (semiaperture 1.4$^\circ$) and $1/100$ (0.6$^\circ$) respectively. The PET
crystal was used to generate 2.0~keV and 4.0~keV polarized radiation, corresponding to the first and
second order of diffraction \citep[$2d=8.742$~\AA{},][]{Henke1993}, while for graphite we exploited
the first three orders at 2.6~keV, 5.2~keV and 7.8~keV \citep[$2d=6.708$~\AA{},][]{Henke1993}. At
3.7~keV we used the prototype source because of the favorable accordance between the aluminum
crystal and the K$\alpha$ line of the calcium tube. In this case the geometry was already
constrained by the use of nearly monochromatic incident photons and hence only a single collimator
(with collimation $1/40$) limited the diffraction angle. Helium flowing in the Bragg diffractometer
was required with PET crystal to avoid severe air absorption of 2.0~keV photons. A diaphragm 2~mm in
diameter was used to illuminate only the small central region of the detector. The setup for each
crystal is summarized in Table~\ref{tab:LineSetup}.

\begin{table}
\centering
\begin{tabular}{cccccc}
\hline
Energy & Crystal & X-ray tube & Collimation & Diaphragm & $\mathcal{P}$ \\
(keV) &  &  &  & mm &  \\
\hline
2.0 & PET, 1st order & \multirow{2}{*}{Ti, 50~W} & \multirow{2}{*}{In: $\frac{1}{40}$, Out:
$\frac{1}{100}$} & \multirow{2}{*}{$\varnothing$~2} & \multirow{2}{*}{$>$0.99} \\
4.0 & PET, 2nd oder &  &  &  &  \\
\\
2.6 & GrD, 1st order & \multirow{3}{*}{Ti, 50~W} & \multirow{2}{*}{In: $\frac{1}{40}$, Out:
$\frac{1}{100}$} & \multirow{3}{*}{$\varnothing$~2} & \multirow{3}{*}{$>$0.99} \\
5.2 & GrD, 2st order &  &  &  &  \\
7.8 & GrD, 3st order &  &  &  &  \\
\\
3.7 & Al, 1st order & Ca, 2~W & Out: $\frac{1}{40}$  & $\varnothing$~2 & 0.9938
\end{tabular}
\caption{Setup for each crystal. ``GrD'' stands for the mosaic graphite grade D crystal.}
\label{tab:LineSetup}
\end{table}

Spectra diffracted by crystals were acquired by means of a Si-PIN Amptek XR100CR spectrometer
with a energy resolution $\sim$200~eV FWHM at 5.9~keV. Lines were fitted with a gaussian profile and
the measured line energy was used to estimate the degree of polarization $\mathcal{P}$ by the
calculation of the diffraction angle and of the interpolated value of $k$. Thanks to the tight
collimation of the two capillary plates on the incident and diffraction directions, the degree of
polarization is always above 99\%. The lack of any significant dilution of polarization is also
confirmed by the narrowness of diffracted lines, which have a FWHM ($\sim$200~eV, 215~eV at 7.8~keV)
consistent with that of fluorescence lines at comparable energies.

\begin{figure}
\centering
\includegraphics[angle=90,width=12cm]{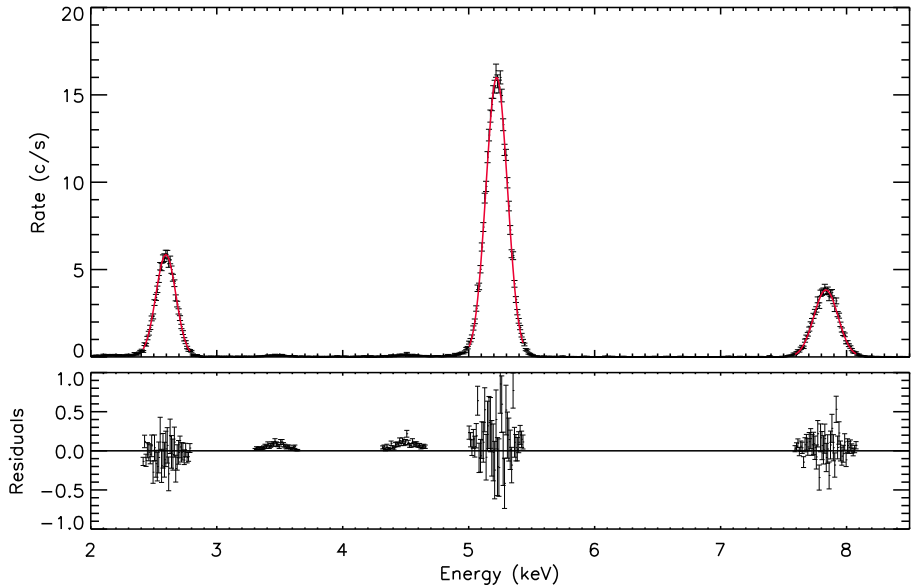}
\caption{Spectrum obtained by diffraction on the graphite crystal of continuum photons produced with
the titanium X-ray tube and acquired with the Amptek XR100CR spectrometer. The first three orders
are visible at 2.6~keV, 5.2~keV and 7.8~keV and gaussian fits of each line are also reported as red
solid lines. The high voltage of the X-ray tube was set to 10~kV to avoid the presence of photons at
energies above the third order. The line at about 3.5~keV is the escape peak of 5.2~keV photons and
that at 4.5~keV is the scattered titanium K$\alpha$.}
\label{fig:GrD_Spectrum}
\end{figure}

As an example, we report in Figure~\ref{fig:GrD_Spectrum} the spectrum for diffraction on the
graphite crystal of radiation generated by the titanium X-ray tube. The first three orders are
prominent at 2.6~keV, 5.2~keV and 7.8~keV but from residuals is also evident a line at about
3.5~keV, which is the escape peak from silicon of the 5.2~keV photons, and a contribution at
4.5~keV. The latter is caused by K$\alpha$ titanium fluorescence photons scattered on the crystal or
on its holder and reaching the detector after passing through the two collimators. By the way, the
two collimators are completely opaque at this energy and then only scatterings at nearly 90$^\circ$
are possible. This implies that even scattered photons are highly polarized perpendicularly to the
scattering plane. Even if the GPD is not able to resolve this line and that at 5.2~keV, the flux of
the former is $<1/300$ than that at 5.2~keV and therefore it will be neglected in the following.

\section{Spectral capabilities}\label{sec:Spectral}

\begin{figure}
\centering
\subfigure[\label{fig:PH_2-4keV}]{\includegraphics[totalheight=4.cm]{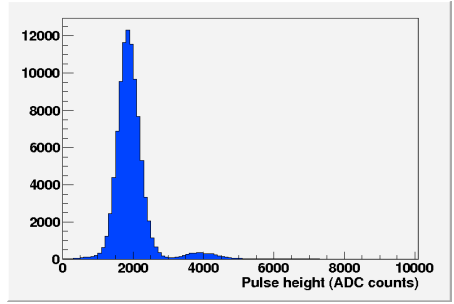}}\hspace{2mm}
\subfigure[\label{fig:PH_2-5-7keV}]{\includegraphics[totalheight=4.cm]{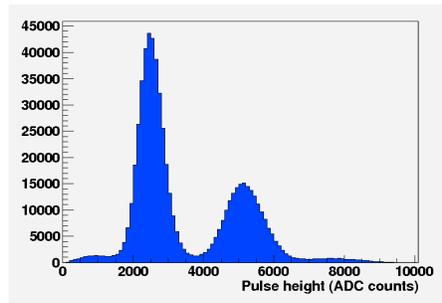}}
\caption{GPD spectrum of monochromatic photons at 2.0~keV and 4.0~keV ({\bf a}) and at 2.6~keV,
5.2~keV and 7.8~keV ({\bf b}).}
\label{fig:PH_all}
\end{figure}

In Figure~\ref{fig:PH_all} we report the spectrum acquired with the GPD for monochromatic (and
polarized) radiation obtained by Bragg diffraction on PET (2.0~keV and 4.0~keV) and graphite
crystals (2.6~keV, 5.2~keV and 7.8~keV). Energy of photons is derived by the sum of the charge
content of hit pixels. The lines are well resolved and this allows us to present in the next Section
the modulation factor for each energy without any significant crosstalk. Lines are fitted with a
Gaussian profile and the results are reported in Table~\ref{tab:LineFit}. Data is filtered (i) to
select spatially the incident beam and (ii) remove the tracks with two or more clusters of hit
pixels, that is we postpone those events ($\lesssim$5\%) which are composed of non-contiguous groups
of pixels to subsequent refined analyses.

\begin{table}
\centering
\begin{tabular}{cccc}
\hline
Energy & E$_{peak}$ & FWHM & $\delta E$/$E$ \\
(keV) & (ADC) & (ADC) & (ADC) \\
\hline
2.0 & 1867 $\pm$ 1 & 689 $\pm$ 3 & 0.369 $\pm$ 0.002 \\
2.6 & 2480 $\pm$ 1 & 797 $\pm$ 3 & 0.321 $\pm$ 0.001 \\
3.7 & 3783 $\pm$ 2 & 1071 $\pm$ 4 & 0.283 $\pm$ 0.001 \\
4.0 & 4013 $\pm$ 3 & 1104 $\pm$ 8 & 0.275 $\pm$ 0.002 \\
5.2 & 5099 $\pm$ 3 & 1248 $\pm$ 8 & 0.245 $\pm$ 0.002 \\
7.8 & 7645 $\pm$ 15 & 1793 $\pm$ 23 & 0.235 $\pm$ 0.003	
\end{tabular}
\caption{Results of the fit to the lines at different energies acquired with the GPD.}
\label{tab:LineFit}
\end{table}

\begin{figure}
\centering
\includegraphics[angle=90,totalheight=5.5cm]{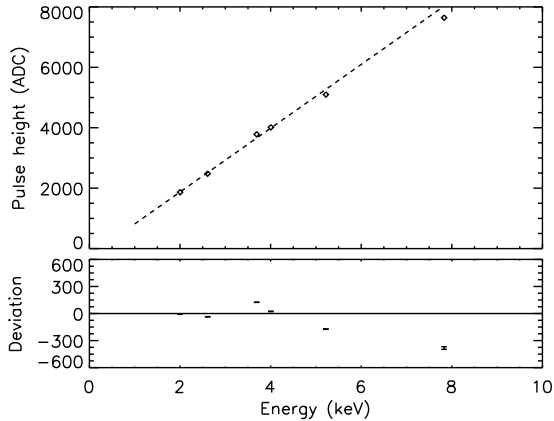}
\caption{Energy calibration of the GPD.}
\label{fig:Cal}
\end{figure}

The relation between the energy and the pulse height, reported in Figure~\ref{fig:Cal}, is
fitted with a line $y = mx+q$, where $q=(-239.7\pm2.3)$~ADC and $m=(1055.8\pm0.8)$~ADC/keV. The
deviations from linearity (5\% at 7.8~keV) can be explained by the insufficient control of high
voltage values and/or the inadequate monitor of the environment conditions. The HV power supply used
(CAEN N470) has a accuracy $\pm$1~V and a long term stability $\pm$2~V, which imply variations on
the gain of the order of $\pm$4\%. The gain $G$ can be normalized by temperature $T$ and pressure
$p$, expressed in Kelvin and torr respectively, with the function:
\begin{equation}
\frac{G^{corr}}{G^{meas}} =
\frac{1}{\exp{\left[C\times\left(\frac{1}{p/T}-\frac{1}{2.533}\right)\right]}}, \label{eq:Gain}
\end{equation}
where $G^{meas}$ and $G^{corr}$ are the measured and corrected values and $C$ is a constant, equal
to 19.1~Torr/K for the RIKEN-80T-LCP GEM \citep{Tamagawa2009}. Changes of temperature cause linear
variations on the gain, $\Delta g \propto -\Delta T$, and fluctuations of 2~K, which are plausible
in our case, results in gain instabilities of 5\%.

\begin{figure}
\centering
\includegraphics[angle=90,totalheight=5.5cm]{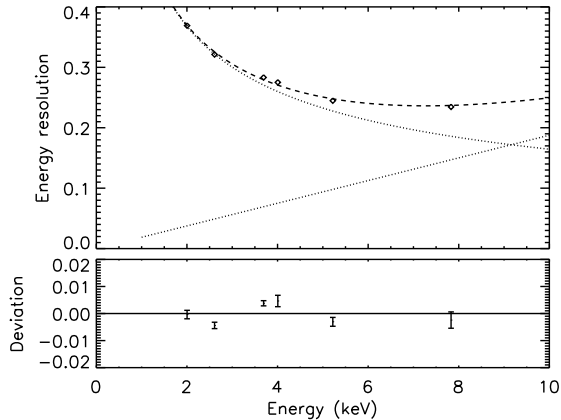}
\caption{Energy resolution as a function of energy. The fit function (dashed line) takes into
account the statistical noise and a linear contribution. These two terms are plotted with dotted
lines.}
\label{fig:EnRes}
\end{figure}

Within Poisson or Poisson-like statistics of ionization and charge amplification \citep{Knoll2000},
the dependence of the energy resolution ${\delta E}/{E}=\mbox{FWHM}/{E_{peak}}$ as a function of
energy is expected to be ${\delta E}/{E}\propto1/\sqrt{E}$ for a gas detector. The relation in the
case of the GPD is reported in Figure~\ref{fig:EnRes}: although the energy resolution decreases with
energy, the sole dependence with $1/\sqrt{E}$ is unsatisfactory. Therefore we used a function:
\begin{equation}
\frac{\delta E}{E} =\sqrt{\left(\frac{k_1}{\sqrt E}\right)^2+\left(k_2\;E\right)^2},
\end{equation}
where $k_1=(0.520\pm0.001)\;\sqrt{\mbox{keV}}$ and $k_2=(0.0188\pm0.0005)$~keV$^{-1}$ (dashed line
in Figure~\ref{fig:EnRes}). The fit function comprises two contributions summed quadratically, the
first being the statistical Poisson noise in the ionization and amplification of the charges. We
used the second term to model in a simple way an additional contribution which seems to weigh more
at higher energy. A significant role within it could be played by incomplete charge collection
due to photoelectron hits on the gas cell surfaces, which depends on the direction of emission but
also on energy through the range (1~mm at 7.8~keV). We positioned the beam in the center of the
detector, but loss of charge can still occur if photons are absorbed near the window or the GEM.
More energetic and hence longer tracks are also spread on larger regions of the GEM and of the ASIC
and then nonuniformities could give a contribution. While the former have been reported at the level
of several percent on spatial scales of a few millimeters \citep{Tamagawa2009}, we are confident
that ASIC contribution is small if not negligible. Pedestals are read-out and substracted
immediately after the event (delay of a few $\mu$s) and pixel noise is very low, 50 electrons ENC
\citep{Bellazzini2006}. Tracks produced by photons at 2.0~keV hit on average 40~pixels (see below)
and then the total contribution of electronic noise is of the order of
$\sqrt{2}\;\sqrt{40}\;50\approx450$~electrons, where the factor $\sqrt{2}$ is for pedestals
substraction. Conversely, 2.0~keV photons produce 71 primary electrons, assuming that the average
energy loss for the creation of a ion-electron pair in DME is 28~eV \citep{Pansky1997}, and Poisson
fluctuations are $\sqrt{f\;71}$, where $f$ is the Fano factor $\sim$0.3 for
DME \citep{Pansky1997}. After the GEM amplification of a factor 500 (see Table~\ref{tab:GPD}),
statistical fluctuations result in 4800~electrons \citep{Knoll2000}, i.e. a factor 10 larger than
electronic noise. Actually, we expect that the main source of the possible electronic contribution
to energy resolution is the spatial nonuniformities in pixels gain, which can't be calibrated
because sufficient precise test capacitances does not allow to be build with the 0.18~mm VLSI
technology.

Spectral performance achieved is overall quite good. The energy resolution at 5.9~keV is less than
24\% and then very close to the requirement of 20\% for space missions that foresee a GPD on-board,
which, however, assume that energy is measured with a dedicated spectroscopic channel taking the
signal from the upper electrode of the GEM. 

\begin{figure}
\centering
\includegraphics[angle=90,totalheight=5cm]{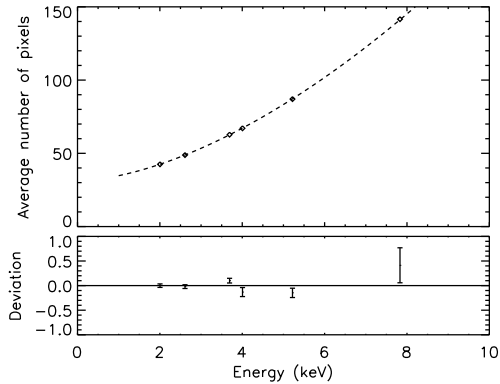}
\caption{Average number of hit pixels per track $p$ as a function of energy. Errors are evaluated as
the ratio between the root mean square of the $p$ distribution and the square root of counts.}
\label{fig:AP}
\end{figure}

Interestingly enough, the average number of hit pixels $\langle p\rangle$ is a very well defined
function of energy (see Figure~\ref{fig:AP}). The dependency can be modelled with a function:
\begin{equation}
\langle p \rangle = k_a+k_b\;\left(\frac{E}{\mbox{1~keV}}\right)^{k_c}, \label{eq:AP}
\end{equation}
with $k_a=(31.15\pm0.13)$, $k_b=(3.58\pm0.05)$ and $k_c=(1.664\pm0.008)$. Note that the
index of the power law is broadly consistent with the energy dependence of electron range in a gas,
reported to be $R=0.71\;E_{\mbox{\scriptsize MeV}}^{1.72}$~g/cm$^2$ under a few hundred of keV by
\citet{Sauli1977}.

In principle the Equation~(\ref{eq:AP}) could be exploited to derive an estimate of the photon
energy, but unfortunately the correlation is rather weak because the number of hit pixels has also a
strong dependence on the photoelectron direction of propagation. Photoelectrons emitted orthogonally
to the collection plane produce smaller tracks than those which instead propagate mostly parallel.
The errors associated to the energy determination by means of the Equation~(\ref{eq:AP}) could be
derived precisely only by the knowledge of the probability function of $p$, but with an approximate
gaussian fit the energy resolution is rather poor ($>60$\%, about constant with energy).

\section{Modulation factor} \label{sec:Mod}

In general, the response of a polarimeter is modulated as a result of the absorption of polarized
photons: the degree and the angle of polarization are derived from the amplitude and the phase of
this modulation, respectively. The modulation curve $\mathcal{M}(\phi)$ of a photoelectric
polarimeter is the number of photoelectrons emitted per each azimuthal angle $\phi$. The direction
of emission is most probably aligned with the electric field of the absorbed photon, the dependency
being expressed by the differential cross section of the interaction \citep{Heitler1954}: 
\begin{equation}
\frac{d\sigma_{ph}^K}{d\Omega}\propto\frac{\sin
^2\theta\cos^2\phi}{\left(1+\beta\cos\theta\right)^4}, \label{eq:PhCrSecDiff}
\end{equation}
where $\beta$ is the photoelectron velocity in units of $c$. $\theta$ and $\phi$ are the
latitudinal and azimuthal emission angles, measured with respect to the direction of
incidence and of polarization respectively. The formula reported above is valid only in
the case of spherically symmetric shells, otherwise corrections which reduce the response to
polarization are required \citep{Ghosh1983}. As far as the GPD is concerned, K-shell is largely the
most probable involved in photoabsorption and then the Equation~(\ref{eq:PhCrSecDiff}) is
sufficiently accurate.

For completely polarized photons, the number of photoelectrons emitted in a certain direction is
modulated as a $\cos^2$ function, while for partially polarized radiation the amplitude is reduced
linearly. Real instruments do not respond perfectly to polarization, in the case of the GPD
because of scatterings with nuclei, gas diffusion and finite size of ASIC pixels. Hence even
for 100\% polarized photons the amplitude of the response, called modulation factor $\mu$, is never
complete but it is calculated as:
\begin{equation}
\mu = \frac{\mathcal{M}_{max}-\mathcal{M}_{min}}{\mathcal{M}_{max}+\mathcal{M}_{min}},
\end{equation}
where $\mathcal{M}_{max}$ and $\mathcal{M}_{min}$ are the maximum and the minimum of the modulation,
which practically is the histogram of photoelectron angles of emission, in the case of completely
polarized photons. Since $\mathcal{M}(\phi)$ is fitted with a function
$\mathcal{M}(\phi)=A+B\cos^2(\phi-\phi_0)$, then:
\begin{equation}
\mu = \frac{B}{2A+B}.
\end{equation}

The modulation factor, and the efficiency $\epsilon$, are the primary parameters used to derive the
sensitivity of a polarimeter. The modulation measured, proportional to $\mu$, is to be compared with
that naturally arising from statistical Poisson fluctuations of the number of photoelectrons
emitted per angular bin, which are inversely proportional to $\sqrt{\epsilon}$ if the background is
negligible as for an experiment at the focus of an X-ray telescope. The product $\mu
\sqrt{\epsilon}$ is called quality factor (see \citet{Muleri2008} for a more extended discussion
on the sensitivity of a polarimeter). Within astrophysical application, the value of $\mu$ is
especially important at low energy, where the largest part of photons are concentrated. Typical
source spectra rapidly decrease with energy, as the efficiency of the instrument and the area of
grazing incidence optics. For a Crab like spectrum between 2~keV and 10~keV and a standard (not
multilayer) telescope, the $\sim$70\% of counts are in the 2-3~keV energy range. For this reason, in
the following we focus our attention on the performance at low energy.

\subsection{Data analysis} \label{sec:Cuts}

The image of the photoelectron track is processed with an algorithm which reconstructs the
absorption point as the center of gravity of the charge distribution (barycenter) and the 
direction of emission as that which maximizes the second moment \citep{Bellazzini2003}. The latter
is basically the direction of elongation of the track. However, for energies higher than
$\sim$3~keV, the initial part of the track with lower ionization density (energy losses are
inversely proportional to the energy) and the point where photoelectron is stopped (Bragg peak) are
resolved. The algorithm selects the pixels at the edge of the track with the lower signal density
and calculates the impact point and the maximum second moment, i.e. the direction of emission of the
photoelectron, considering only these pixels. The initial direction of photoelectrons is used to
create the histogram of emission angles, that is the modulation curve, which shows a $\cos^2$
modulation if the absorbed photons are polarized while it is flat for unpolarized radiation.
Depending on the polarization degree to be detected, the modulation curve must be constructed with a
large number of photons, at least several tens of thousands, to achieve a result which is
statistically significant.

\begin{figure}
\centering
\subfigure[\label{fig:Track_2keV}]{\includegraphics[totalheight=5.2cm]{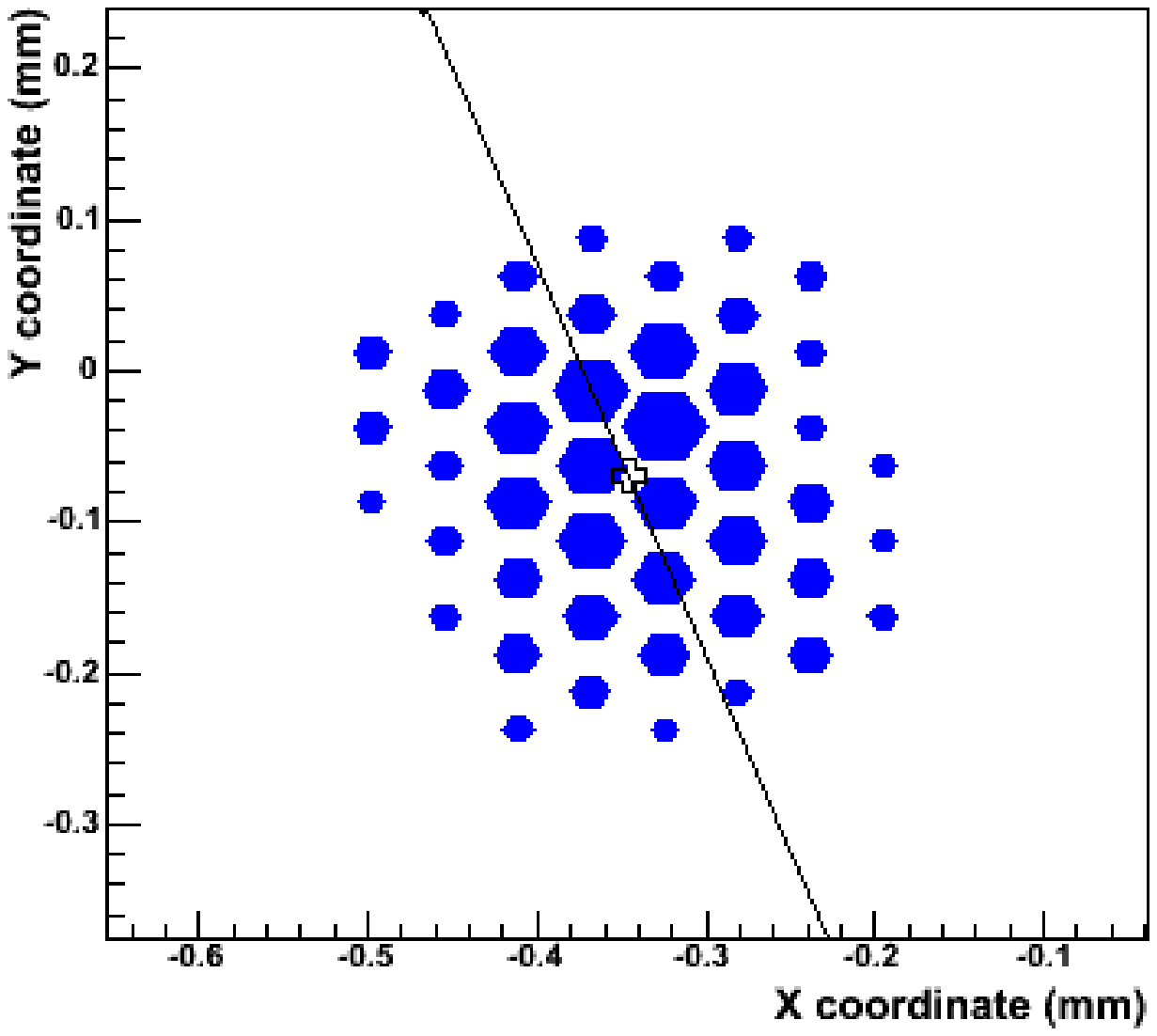}}\hspace{2mm}
\subfigure[\label{fig:Track_5keV}]{\includegraphics[totalheight=5.2cm]{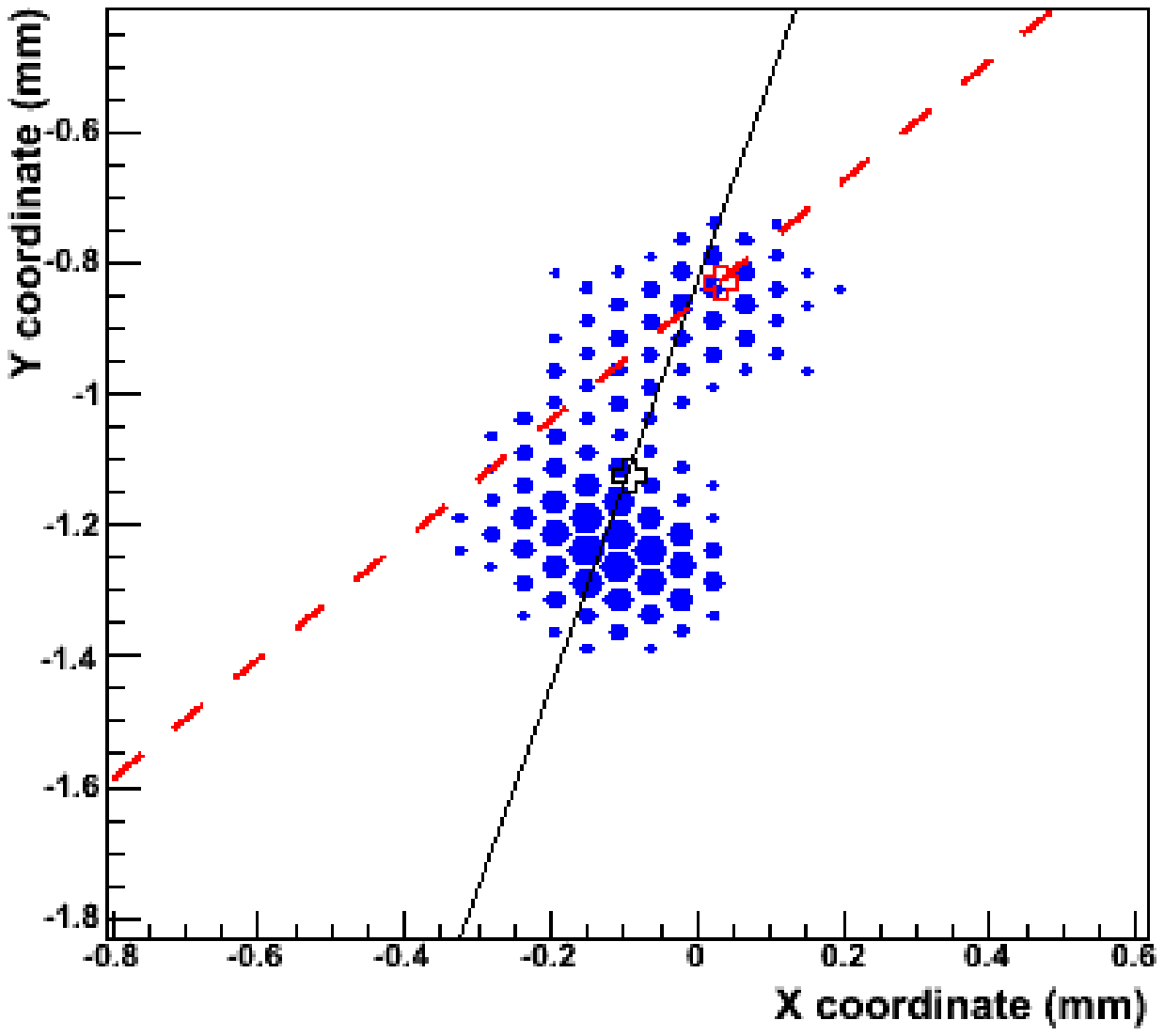}}
\caption{Examples of tracks at 2.0~keV ({\bf a}) and 5.2~keV ({\bf b}). The reconstructed direction
of emission obtained by analyzing the entire track and resolving the first part are the black
solid line and the red dashed one, respectively. The barycenter and the impact point are the black
and the red crosses.}
\label{fig:Tracks}
\end{figure}

In Figure~\ref{fig:Tracks} we report two examples of tracks at 2.0~keV and 5.2~keV to show the
effectiveness of our reconstruction algorithm. At low energy the track is so short that any
substructure is blurred by diffusion during the drift in the gas cell. However a fraction of the
original polarimetric information is still present in the elongation of the charge distribution,
which is correlated with the initial direction of photoelectron emission as demonstrated by the
measurement of a modulation factor value different from zero (see below). Instead, when tracks are
more energetic and hence longer, the final part is clearly distinguished by the denser charge
density. The direction reconstructed at the second step, reported as the red dashed line in
Figure~\ref{fig:Track_5keV}, is evidently a better approximation of the actual direction of emission
than that obtained at the first step as the elongation of the charge distribution (black solid line
in Figure). As a matter of fact, the improvement of the modulation factor, that is the response to
polarization, achieved passing from the first to second step is 28\% at 5.2~keV and 50\% at 7.8~keV.

Beyond selecting spatially the incident beam and excluding the tracks with more than one cluster,
as explained in Section~\ref{sec:Spectral}, we remove from the analysis on modulation factor those
events which (i) hit less than 27 pixels and (ii) have an asymmetry less than a threshold
that is different for each energy. The first cut is directly related to the GEM used, which has a
pitch larger than usual (80~$\mu$m). When primary electrons are amplified by the GEM, the charges
are constrained in the holes and this actually samples the event on the microscopic structure of the
GEM. If the track is so short to involve only a few holes, the amplified charge distribution
reflects significantly the GEM pattern. This is hexagonal and then produces on the elongation of
the track, ultimately related to the reconstructed direction of emission, a systematic effect with
periodicity 60$^\circ$, namely the amplification makes most probable the directions aligned with GEM
holes, which correspond to integer multiples of 60$^\circ$. Pixels on the ASIC are arranged more
finely than holes in the GEM (pitch 50~$\mu$m vs 80~$\mu$m) and removing those events which hit
less than 27 pixels implies that we exclude from analysis tracks passed through only 10 GEM holes or
less.

\begin{figure}
\centering
\includegraphics[totalheight=5cm]{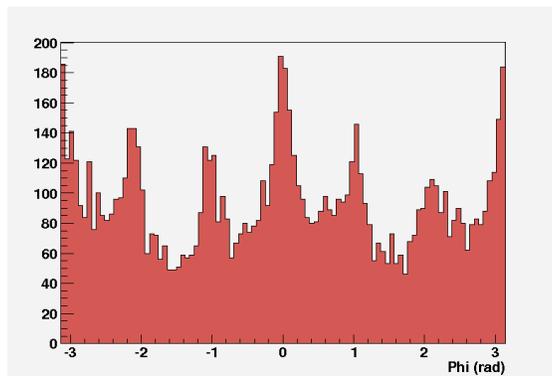}
\caption{Systematic effects which emerge by selecting only tracks which hit less than 27~pixels at
2.0~keV. The 60$^\circ$ periodicity reflects the hexagonal pattern of the GEM, whose pitch is
larger than usual.}
\label{fig:Theta0_27}
\end{figure}

The modulation curve for events which hit less than 27~pixels is reported in
Figure~\ref{fig:Theta0_27}. The sharp peaks clearly mirror the hexagonal pattern of the GEM and
significantly affect the total modulation curve because the fraction of events with less than
27~pixels is 10\% at 2.0~keV (7.6\% at 2.6~keV). 

The asymmetry of a track (or eccentricity $e$) is defined as the ratio between the maximum and
minimum values of the second moment of the charge distributions $M^{II}$,
$e={M_{max}^{II}}/{M_{min}}^{II}$. A higher value of $e$ means that the track is more developed in a
certain direction and then the information on the initial direction of emission is more preserved.
On the contrary, almost round tracks ($e\approx1$) have no preferred directions and then the large
part of polarimetric information is lost.

We used the second cut on eccentricity to make our results comparable to what previously presented
for the GPD filled with helium and DME by \citet{Muleri2008}. These authors reported the modulation
factor with different data selections aimed to optimize the quality factor $\mu\sqrt{\epsilon}$,
and, from that analysis, best results were obtained by removing about 25\% of events. In this paper
we follow a specular approach as for previous measurements, we change the cut on the asymmetry of
tracks to exclude from analysis about 25\% of tracks to maximize the quality factor.

\subsection{Comparison to Monte Carlo results and other mixtures}

\begin{figure}
\centering
\includegraphics[totalheight=5cm]{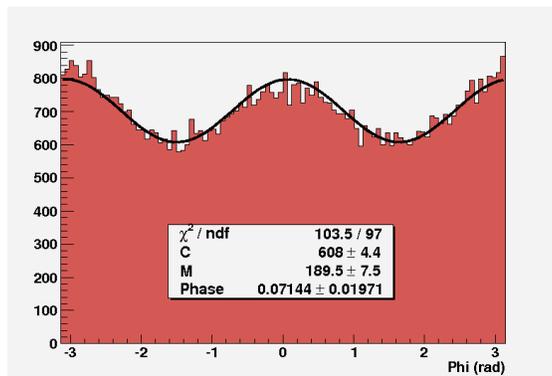}
\caption{Modulation measured for 2.0~keV polarized photons.}
\label{fig:Theta0_2keV}
\end{figure}

An example of the modulation curve obtained at 2.0~keV, particularly significative because it refers
to the lowest energy at which a photoelectric polarimeter has ever been tested, is reported in
Figure~\ref{fig:Theta0_2keV}. At this energy the modulation factor is 13.5\%, while the values
measured at other energies are reported in Table~\ref{tab:ModFit}.

\begin{table}
\centering
\begin{tabular}{cccccc}
\hline
Energy & $\mu$   & Cuts & \textsc{FdR} & $\chi_\nu^2$ & Counts \\
(keV) &    &  & (\%) & \\
\hline
2.0 & 0.135 $\pm$ 0.005 & \#$_p>$27, $e>$1.18 & 23.5 & 1.1 & 70379 \\
2.6 & 0.267 $\pm$ 0.004 & \#$_p>$27, $e>$1.20 & 22.0 & 1.6 & 91055 \\
3.7 & 0.402 $\pm$ 0.003 & \#$_p>$27, $e>$1.39 & 24.4 & 1.3 & 88968 \\
4.0 & 0.426 $\pm$ 0.006 & \#$_p>$27, $e>$1.48 & 24.4 & 1.2 & 27279 \\
5.2 & 0.496 $\pm$ 0.005 & \#$_p>$27, $e>$2.00 & 23.8 & 0.8 & 46141 \\
7.8 & 0.589 $\pm$ 0.010 & \#$_p>$27, $e>$3.70 & 24.4 & 0.9 & 9112
\end{tabular}
\caption{Modulation factor measured at different energies. Cuts are applied on the
number of hit pixels (\#$_p$) and on the eccentricity of tracks $e$. The fraction of data removed
(\textsc{FdR}), the reduced $\chi_\nu^2$ of the $\cos^2$ fit (97 d.o.f.) and the
counts after cuts are also reported.}
\label{tab:ModFit}
\end{table}

\begin{figure}
\centering
\includegraphics[angle=90,totalheight=6cm]{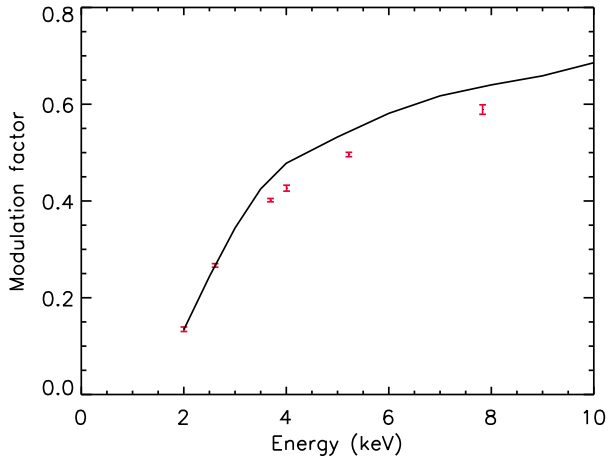}
\caption{Measured values of the modulation factor (red points) compared with Monte Carlo results
(solid line).}
\label{fig:ModFac}
\end{figure}

Measured modulation factor is compared to the estimates derived from Monte Carlo simulations in
Figure~\ref{fig:ModFac}. Noteworthily we achieved expected performance at low energy (even if with
modest cuts), while at higher energies measured values are systematically lower than expected.
This discrepancy is quite small, $\lesssim$10\%, and actually does not impact on the sensitivity of
the instrument because emission from astrophysical sources is strongly concentrated at low energy
where expected performance is entirely confirmed. Nonetheless this issue is probably a signature
that Monte Carlo can still be refined and the discrepancy will be investigated to exploit in the
best way the whole energy range of the instrument. The measurement of possible variations of
polarization signature with energy is at the basis of the success of a X-ray polarimetry mission and
the exploitation of the largest energy range possible, even better if it extends the ``classical''
2-10~keV energy interval, is mandatory.

Finally it is useful to compare the modulation factor of the GPD presented in this paper with
that already reported by \citet{Muleri2008} for a mixture composed of 20\% helium and 80\% DME at
1~atm. Despite the fact that the partial pressure of DME is (almost) the same for the two gases,
there are some differences. While at lower energies the behavior is quite similar, the DME mixture
apparently is less effective at high energy and at 5.2~keV the quality factor is $\sim$10\% lower.
Hence the presence of a small fraction of helium seems to favor reconstruction of longer tracks.

\begin{table}
\centering
\begin{tabular}{ccccc}
& \multicolumn{2}{c}{{\bf DME}, this work} & \multicolumn{2}{c}{\bf He-DME} \\
\hline
Energy & $\mu$ & $\mu\sqrt{\epsilon}$ & $\mu$ & $\mu\sqrt{\epsilon}$ \\
(keV) &  &  &  &  \\
\hline
2.0 & 0.135 $\pm$ 0.005 & 0.058 $\pm$ 0.002 & ---   & ---   \\
2.6 & 0.267 $\pm$ 0.004 & 0.103 $\pm$ 0.001 & 0.276 $\pm$ 0.014 & 0.101 $\pm$ 0.005 \\
3.7 & 0.402 $\pm$ 0.003 & 0.105 $\pm$ 0.001 & 0.431 $\pm$ 0.012 & 0.107 $\pm$ 0.003 \\
4.0 & 0.426 $\pm$ 0.006 & 0.101 $\pm$ 0.001 & ---   & ---   \\
5.2 & 0.496 $\pm$ 0.005 & 0.082 $\pm$ 0.001 & 0.545 $\pm$ 0.010 & 0.091 $\pm$ 0.002 \\
7.8 & 0.589 $\pm$ 0.010 & 0.053 $\pm$ 0.001 & ---   & ---
\end{tabular}
\caption{Comparison of the modulation factor $\mu$ and quality factor $\mu\sqrt{\epsilon}$
measured for DME at 0.79~atm and a mixture He 20\% and DME 80\% at 1~atm presented by
\citet{Muleri2008}.}
\label{tab:ModComp}
\end{table}

\section{Conclusion}

The characterization of the GPD presented in this paper represents one step forward to the
optimization of the instrument as a photoelectric polarimeter. In place of the standard mixture
composed of a small fraction of helium (20\%) and DME (80\%) at 1~atm, we tested the GPD filled with
pure DME at 0.79~atm. Measured modulation factor successfully confirms Monte Carlo simulations at
lower energy, while above 3.7~keV there is a discrepancy $\lesssim$10\% with respect both the
expected value and the sensitivity measured for the (similar) standard mixture. Although this issue
does not significantly affect the performance of the GPD, which are mostly determined at low energy,
it deserves further investigations to exploit in the best way the whole energy range of the
instrument.
Conversely the modulation factor measured at 2.0~keV, the lowest energy ever presented for a
photoelectric polarimeter, is relatively large, 13.5\%, and encourages attempts to reduce the
threshold of the instrument below this energy value. This could allow to address interesting
scientific objectives, like the study of the (possibly) highly polarized thermal emission from the
surface of cooling neutron stars.

We also discussed, for the first time in a systematic way, the spectral capabilities of the GPD. 
The spectrum of absorbed photons is obtained by summing the charge content of hit pixels and, in
principle, it could be heavily affected by nonuniformities in the response of the pixels. As a
matter of fact, the energy resolution is quite good, 24\% at 5.9~keV, that is very close to that of
standard proportional counters and to the requirement (20\%) defined for future space missions that
include the GPD. The dependency with energy, beyond a $1/\sqrt{E}$ contribution that is naturally
explained as due to Poisson fluctuations in production and amplification of primary charges, shows
an additional term whose origin will be subject of further investigation.

In conclusion, DME appears as an interesting alternative to mixtures of helium and DME. Despite the
small reduction of the modulation factor at high energy, performance at low energy is confirmed.
Possibly even more important, DME provides very good spectral capabilities, which are a mandatory
complement of the polarimetric sensitivity to pursue the scientific objectives of any X-ray
polarimetry mission.

\section*{Acknowledgments} 

The activity was supported by ASI contracts I/088/060 and I/012/08/0.

\bibliography{References}
\bibliographystyle{elsarticle-harv}

\end{document}